\documentclass[prb,twocolumn,aps,amssymb]{revtex4}
\usepackage{graphicx}
\begin{document}

\title{Critical Entropy of Quantum Heisenberg Magnets on Simple-Cubic Lattices}
\author{Stefan Wessel}
\affiliation{Institut f\"ur Theoretische Physik III,
Universit\"at Stuttgart, Pfaffenwaldring 57, 70550 Stuttgart, Germany}

\begin{abstract}
We analyze the temperature dependence of the entropy of the spin-1/2 Heisenberg model on the three-dimensional simple-cubic 
lattice, for both the case of antiferromagnetic and ferromagnetic nearest neighbor exchange interactions. 
Using optimized extended ensemble quantum Monte Carlo simulations, we extract the entropy at the critical temperature for 
magnetic order from a finite-size scaling analysis. 
For the antiferromagnetic case, the critical entropy density equals $0.341(5)k_B$, whereas for the 
ferromagnet, a larger value of $0.401(5)k_B$ is obtained. We compare our simulation results to estimates 
put forward recently in studies  assessing  means of 
realizing the antiferromagnetic N\'eel state in ultra-cold fermion gases in optical lattices. 
\end{abstract}
\maketitle
\section{Introduction}
Ultra-cold atom gases provide a unique opportunity to study basic models of strongly interacting many-body systems in a highly controlled experimental framework~\cite{ bloch08}. After the pioneering  realization of the superfluid-to-Mott insulator transition of ultra-cold bosonic atoms~\cite{greiner02}, recently different groups reported the realization of a Mott-insulating state also for ultra-cold ${}^{40}$K fermionic atoms on three-dimensional simple-cubic lattices~\cite{joerdens08,schneider08a}. A  next major step would be the realization and identification of a low-temperature antiferromagnetically ordered N\'eel state in such systems.
In fact, various recent studies explored the prospects of realizing the antiferromagnetic N\'eel state in an 
ultra-cold gas of fermionic atoms confined to an optical lattice~\cite{werner05,dare07,deleo08,koetsier08,ho09,bernier09,mathy09,gottwald09}. 
A  quantity that is important in assessing the relevant temperature scale is the upper bound on the entropy allowed in order to transfer the atoms into the antiferromagnetically ordered state. For a three-dimensional fermionic Hubbard model on a simple-cubic lattice at half-filling, estimates of the entropy $S$ as a function of temperature, as well as 
the ratio $U/t$ between the onsite repulsion $U$ and the hopping amplitude $t$, were obtained within a single-site dynamical mean-field theory (DMFT) approach~\cite{werner05}. As reported in Ref.~\onlinecite{werner05},  DMFT however overestimates the N\'eel temperature by about $50\%$ in the intermediate coupling regime, as compared to direct quantum Monte Carlo simulations based on systems with up to $10^3$ lattice sites~\cite{staudt00}.  Obtaining in a similar way the entropy from direct quantum Monte Carlo simulations is challenging, and typically involves  integration of the specific heat over an extended temperature range.

However, in the limit of large $U/t$, the spin physics of the Hubbard model is well known to be described by a nearest-neighbor spin-1/2 quantum Heisenberg model, with an antiferromagnetic exchange coupling $J=4 t^2/U>0$, obtained within perturbation theory around the strong coupling limit $t=0$. 
This model is accessible to large scale quantum Monte Carlo simulations and moreover -- as shown below -- it is possible to provide precise values of the entropy in this large-$U$ limit. In particular, one can obtain the value of the critical entropy density (i.e. the entropy per lattice site) $S_C$, below which antiferromagnetic order emerges. 
Nevertheless, thus far no systematic quantum Monte Carlo study concerning the entropy has been reported. On the other hand, different  estimates of $S_C$ for the Heisenberg antiferromagnet have been put forward recently. In Ref.~\onlinecite{werner05}, a Schwinger boson approach~\cite{aua} was employed, leading to the estimate that $S_C$ is about $50\%$ of the mean-field value $S_C^{MF}=k_B \ln (2)$. A more recent study~\cite{koetsier08} reported a value of $S_C$ obtained from a fluctuation-corrected mean-field approach, leading to a reduction in $S_C$ of only $20\%$ from the mean-field value $S_C^{MF}$. This rather large discrepancy among current estimates of $S_C$ calls for a clarification based on numerically exact 
simulations of the Heisenberg model. 
\begin{figure}[t]
\begin{center}
\includegraphics[width=8.5cm]{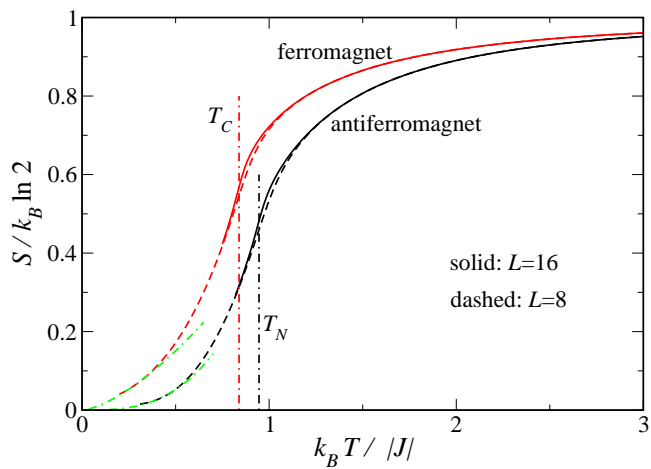}
\caption{(Color online) Temperature dependence of the entropy of the spin-1/2 quantum Heisenberg antiferromagnet and ferromagnet on the simple-cubic lattice. For both cases, results from quantum Monte Carlo simulations are shown for finite systems with $L=8$ (dashed lines) and $L=16$ (solid lines). Error bars for the shown data are below the line thickness. 
Vertical lines show the transition temperatures $T_N$ and $T_C$ for the antiferromagnet and the ferromagnet, respectively. 
Green dashed-dotted curves denote  fits to the algebraic low-$T$ scaling $S\propto T^3$ ($S\propto T^{3/2}$) for the antiferromagnet (ferromagnet).
}
\end{center}
\end{figure}
Here, we present an unbiased estimate of the temperature dependence of the entropy and in particular the critical entropy density by combining extended ensemble quantum Monte Carlo simulations with a finite-size scaling analysis based on the known critical properties of the quantum Heisenberg model. 
For comparison, we consider in the following both the ferromagnetic and the antiferromagnetic spin-1/2 Heisenberg model on the three-dimensional simple-cubic lattice. These models are described by a Hamiltonian
\begin{equation}
 H=J\sum_{i,j} \mathbf{S}_i\cdot \mathbf{S}_j, 
\end{equation}
where $\mathbf{S}_i$ denotes a spin-1/2 degree of freedom on lattice site $i$, and $J$ is nearest-neighbor exchange interaction, with $J>0$ ($J<0$),  for the antiferromagnetic (ferromagnetic) case. From previous studies,  the locations of the transition temperatures to the ordered phases have been determined as $k_B T_N/J=0.946(1)$~\cite{sandvik98a} (N\'eel temperature) for the antiferromagnet, and $k_B T_C/|J|=0.839(1)$~\cite{troyer04a} (Curie temperature) for the ferromagnet, based on quantum Monte Carlo simulations. 

\section{Method}
In order to extract the temperature dependence of the entropy, we use an optimized extended ensemble approach~\cite{trebst04,wessel07a}, that is based on a generalization of the Wang-Landau~\cite{wang01} algorithm to the case of quantum Monte Carlo simulations~\cite{troyer03a,troyer04a}, performed within the stochastic series expansion representation~\cite{sandvik99b} using multi-cluster deterministic loop updates~\cite{ henelius00}. Within this approach, one obtains Monte Carlo estimates of the expansion coefficients $g(n)$ of the high-temperature series expansion of the partition function $Z$ in the inverse temperature $\beta=1/(k_B T)$,
\begin{equation}
 Z=\mathrm{Tr}e^{-\beta H}=\sum_{n\geq0} g(n) \beta^n,
\end{equation}
for a given system of  $N=L^3$ lattice sites. Here, $L$ denotes the linear size of the simulated finite cube, and we employ periodic boundary conditions in all directions. From the expansion coefficients $g(n)$, the free energy 
\begin{equation}
 F=-\frac{1}{\beta}\ln Z=-\frac{1}{\beta}\ln\sum_{n\geq0} g(n) \beta^n,
\end{equation}
the internal energy 
\begin{equation}
 E=\frac{1}{Z}\sum_{n\geq0} n \: g(n) \beta^{n-1},
\end{equation}
and  the entropy 
\begin{equation}
 S=\frac{E-F}{T}
\end{equation}
are obtained as
continuous functions of the temperature $T$. In practice, as discussed in Ref.~\onlinecite{troyer03a}, we obtain the expansion coefficients  up to a upper cutoff $\Lambda$, that is chosen sufficiently large in order to reliably calculate the thermodynamic properties of the finite system based on the coefficients $g(n)$, $n=0,...,\Lambda$ down to a given temperature scale $\beta_\Lambda$. The required $\Lambda$ scales linear in both $\beta_\Lambda$ and the system size $N$. In our simulations, we consider finite systems up to $L=20$, for which $\Lambda=16000$ is required in order to obtain  thermodynamic properties down to a temperature  of about $k_B T \approx 0.7|J|$. Repeated simulations with independent random streams results in statistical error bars on the thermodynamic quantities.
\begin{figure}[t]
\begin{center}
\includegraphics[width=8.5cm]{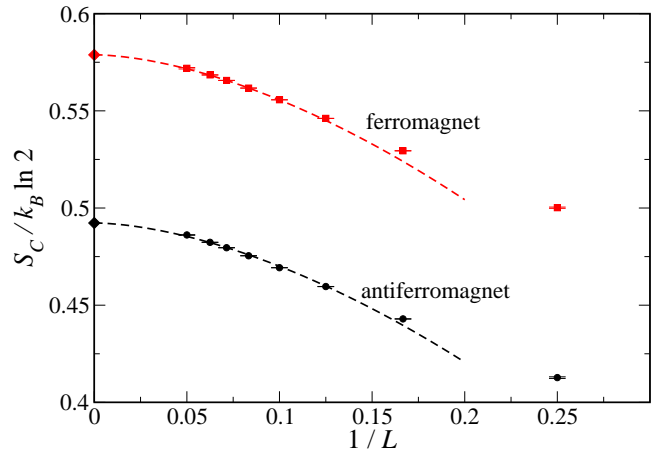}
\caption{(Color online) Finite-size scaling of the critical entropy density $S_C(L)$ with the inverse linear system size $1/L$ for the spin-1/2 quantum Heisenberg antiferromagnet and ferromagnet on the simple-cubic lattice. The quantum Monte Carlo data are shown by symbols, and the dashed lines represent fits of the finite-size data for $L>6$ to the finite-size scaling form in Eq.~(6). The results for $S_C$ in the thermodynamic limit are indicated by symbols as well. 
}
\end{center}
\end{figure}

\section{Results}
In Fig.~1, we present the results for the entropy from our simulations for  both the antiferromagnetic and the ferromagnetic Heisenberg model. For both cases, we show data for  systems with $L=8$ and $L=16$. Outside a narrow range around the transition temperature, the entropy curves obtained for the two different system sizes compare well, and thus essentially represent the behavior of the entropy density in the thermodynamic limit. Also shown in Fig.~1 are fits of the numerical data to the low-$T$  scaling $S\propto T^3$ and  $S\propto T^{3/2}$ for the antiferromagnet and ferromagnet, respectively. This difference in the low-$T$ scaling behavior relates to the linear (quadratic) dispersion of the magnetic spin-wave excitations in the antiferromagnetic (ferromagnetic) case~\cite{aua}. 

However, in order to extract the critical entropy $S_C$ in the thermodynamic limit, we need to perform a finite-size scaling analysis of the finite-size data $S_C(L)$ taken at the transition temperature. For this purpose, we use a finite-size scaling ansatz
\begin{equation}
 S_C-S_C(L)\propto L^{1/\nu-d},
\end{equation}
where $\nu$ is the critical exponent of the correlation length, and $d$ the dimensionality of the system. 
Starting from $S=-\partial F / \partial T$, 
this relation follows from the scaling behavior of the free energy $F\propto \xi^{-d}$ with the correlation length $\xi$, that scales as $\xi\propto |t|^{-\nu}$ when the reduced temperature $t\rightarrow 0$, by the usual finite-size scaling ansatz, that at criticality $\xi$ relates to the linear system size $L$. 
For the current case of a three-dimensional system ($d=3$) with $O(3)$ symmetry, $\nu=0.7112(5)$~\cite{campostrini02}.
Figure~2 shows that the finite-size data indeed fits well to the above scaling form in the range of system sizes $L>6$.
From the extrapolated values, we obtain $S_C=0.341(5)k_B=0.492(7) S_C^{MF}$ for the antiferromagntic model,  whereas for the 
ferromagnet, a larger value of $0.401(5)k_B=0.578(7)S_C^{MF}$ results. We also performed an algebraic fit without biasing the exponent, which returned identical values of $S_C$ within the statistical error bars. 

While the ferromagnet has a critical temperature $T_C$ that is about $11\%$ \textit{lower} than $T_N$, it orders already below an entropy that is about $17\%$ \textit{higher} than the critical entropy of the antiferromagnet. The higher critical entropy directly exhibits  a reduced efficiency of thermal disorder in destroying long-range magnetic order.  The lower transition temperature of the ferromagnetic is due to a significantly enhanced production of entropy upon heating up the system, which follows  from both Fig.~1, as well as from the mentioned difference in the low-$T$ behavior of the entropy. 

\section{Conclusions}
Based on numerically exact quantum Monte Carlo simulations, we obtained the temperature dependence of the entropy for both the antiferromagnetic and the ferromagnetic spin-1/2 Heisenberg model on the simple-cubic lattice. Focusing on the region close to the magnetic ordering transition, we employed a finite-size scaling analysis to extract the critical entropy density below which long-ranged magnetic order sets in. In case of the antiferromagnetic model, our result for the critical entropy is consistent with a $50\%$ reduction from the mean-field value, that was estimated recently from a Schwinger boson calculation\cite{werner05}. This clarifies among the different values of the critical entropy reported previously\cite{werner05,koetsier08}.
In comparison, we found  the ferromagnet to be more robust to thermal disorder, which at criticality is about $17\%$
larger  for the ferromagnet than for the antiferromagnet in terms of the critical entropy. 
We consider the current results for the temperature dependence of the magnetic entropy of value also for future calculations of the entropy for the full fermionic Hubbard model beyond the strongly interacting regime. 

\begin{acknowledgements}
We thank N. Bl\"umer, A. Muramatsu, and L. Pollet
for discussions, and
acknowledge the allocation of CPU time on the HLRS Stuttgart 
and NIC J\"ulich supercomputers. Support was also provided through the DFG within SFB/TRR 21.
\end{acknowledgements}

\end{document}